%

\input vanilla.sty


\font\ninerm=cmr9
\font\sevenrm=cmr7
\font\sixrm=cmr6
\font\fiverm=cmr5
\font\ninei=cmmi9
\font\sixi=cmmi6
\font\fivei=cmmi6
\font\ninesy=cmsy9
\font\sixsy=cmsy6
\font\fivesy=cmsy5
\font\tenex=cmex10
\font\nineit=cmti9
\font\ninesl=cmsl9
\font\ninett=cmtt9
\font\ninebf=cmbx9
\font\sixbf=cmbx6
\font\fivebf=cmbx5

\def\ninepoint{\def\rm{\fam0\ninerm}
  \textfont0=\ninerm \scriptfont0=\sixrm \scriptscriptfont0=\fiverm
  \textfont1=\ninei \scriptfont1=\sixi \scriptscriptfont0=\fivei
  \textfont2=\ninesy \scriptfont2=\sixsy \scriptscriptfont2=\fivesy
  \textfont3=\tenex \scriptfont3=\tenex \scriptscriptfont3=\tenex
  \textfont\itfam=\nineit  \def\it{\fam\itfam\nineit}%
  \textfont\slfam=\ninesl  \def\sl{\fam\slfam\ninesl}%
  \textfont\ttfam=\ninett  \def\tt{\fam\ttfam\ninett}%
  \textfont\bffam=\ninebf  \scriptfont\bffam=\sixbf
   \scriptscriptfont\bffam=\fivebf  \def\bf{\fam\bffam\ninebf}%
  \normalbaselineskip=11pt
  \setbox\strutbox=\hbox{\vrule height8pt depth3pt width0pt}%
  \let\sc=\sevenrm  \normalbaselines\rm}

\font\eightrm=cmr8
\font\sevenrm=cmr7
\font\sixrm=cmr6
\font\fiverm=cmr5
\font\eighti=cmmi8
\font\sixi=cmmi6
\font\fivei=cmmi6
\font\eightsy=cmsy8
\font\sixsy=cmsy6
\font\fivesy=cmsy5
\font\tenex=cmex10
\font\eightit=cmti8
\font\eightsl=cmsl8
\font\eighttt=cmtt8
\font\eightbf=cmbx8
\font\sixbf=cmbx6
\font\fivebf=cmbx5

\def\eightpoint{\def\rm{\fam0\eightrm}
  \textfont0=\eightrm \scriptfont0=\sixrm \scriptscriptfont0=\fiverm
  \textfont1=\eighti \scriptfont1=\sixi \scriptscriptfont0=\fivei
  \textfont2=\eightsy \scriptfont2=\sixsy \scriptscriptfont2=\fivesy
  \textfont3=\tenex \scriptfont3=\tenex \scriptscriptfont3=\tenex
  \textfont\itfam=\eightit  \def\it{\fam\itfam\eightit}%
  \textfont\slfam=\eightsl  \def\sl{\fam\slfam\eightsl}%
  \textfont\ttfam=\eighttt  \def\tt{\fam\ttfam\eighttt}%
  \textfont\bffam=\eightbf  \scriptfont\bffam=\sixbf
   \scriptscriptfont\bffam=\fivebf  \def\bf{\fam\bffam\eightbf}%
  \normalbaselineskip=9pt
  \setbox\strutbox=\hbox{\vrule height7pt depth2pt width0pt}%
  \let\sc=\sixrm  \normalbaselines\rm}

\catcode `\!=11
\catcode `\@=11

 



\let\!tacr=\\ 


\newdimen\LineThicknessUnit 
\newdimen\StrutUnit            
\newskip \InterColumnSpaceUnit  
\newdimen\ColumnWidthUnit     
\newdimen\KernUnit

\let\!taLTU=\LineThicknessUnit 
\let\!taCWU=\ColumnWidthUnit   
\let\!taKU =\KernUnit          

\newtoks\NormalTLTU
\newtoks\NormalTSU
\newtoks\NormalTICSU
\newtoks\NormalTCWU
\newtoks\NormalTKU

\NormalTLTU={1in \divide \LineThicknessUnit by 300 }
\NormalTSU ={\normalbaselineskip
  \divide \StrutUnit by 11 }  
\NormalTICSU={.5em plus 1fil minus .25em}  
\NormalTCWU ={.5em}
\NormalTKU  ={.5em}

\def\NormalTableUnits{%
  \LineThicknessUnit   =\the\NormalTLTU
  \StrutUnit           =\the\NormalTSU
  \InterColumnSpaceUnit=\the\NormalTICSU
  \ColumnWidthUnit     =\the\NormalTCWU
  \KernUnit            =\the\NormalTKU}
 
\NormalTableUnits



\newcount\LineThicknessFactor    
\newcount\StrutHeightFactor      
\newcount\StrutDepthFactor       
\newcount\InterColumnSpaceFactor 
\newcount\ColumnWidthFactor      
\newcount\KernFactor
\newcount\VspaceFactor

\LineThicknessFactor    =2
\StrutHeightFactor      =8
\StrutDepthFactor       =3
\InterColumnSpaceFactor =3
\ColumnWidthFactor      =10
\KernFactor             =1
\VspaceFactor           =2


\newcount\TracingKeys 
\newcount\TracingFormats  


\def\BeginTableParBox#1{%
  \vtop\bgroup 
    \hsize=#1
    \normalbaselines 
    \let~=\!ttTie
    \let\-=\!ttDH
    \the\EveryTableParBox} 
  
\def\EndTableParBox{%
    \MakeStrut{0pt}{\StrutDepthFactor\StrutUnit}
  \egroup} 

\newtoks\EveryTableParBox
\EveryTableParBox={%
  \parindent=0pt
  \raggedright
  \rightskip=0pt plus 4em 
  \relax}


\newtoks\EveryTable
\newtoks\!taTableSpread


\newskip\LeftTabskip
\newskip\RightTabskip


\newcount\!taCountA
\newcount\!taColumnNumber
\newcount\!taRecursionLevel 

\newdimen\!taDimenA  
\newdimen\!taDimenB  
\newdimen\!taDimenC  
\newdimen\!taMinimumColumnWidth

\newtoks\!taToksA

\newtoks\!taPreamble
\newtoks\!taDataColumnTemplate
\newtoks\!taRuleColumnTemplate
\newtoks\!taOldRuleColumnTemplate
\newtoks\!taLeftGlue
\newtoks\!taRightGlue

\newskip\!taLastRegularTabskip

\newif\if!taDigit
\newif\if!taBeginFormat
\newif\if!taOnceOnlyTabskip



\def\TaBlE{%
  T\kern-.27em\lower.5ex\hbox{A}\kern-.18em B\kern-.1em
    \lower.5ex\hbox{L}\kern-.075em E}



{\catcode`\|=13 \catcode`\"=13
  \gdef\ActivateBarAndQuote{%
    \ifnum \catcode`\|=13
    \else
      \catcode`\|=13
      \def|{%
        \ifmmode
          \vert
        \else
          \char`\|
        \fi}%
    \fi
    \ifnum \catcode`\"=13
    \else
      \catcode`\"=13
      \def"{\char`\"}%
    \fi}}
 
{\catcode `\|=12 \catcode `\"=12 

}


\def\!thMessage#1{\immediate\write16{#1}\ignorespaces}
 
\let\!thx=\expandafter

\def\!thGobble#1{} 

\def\\{\let\!thSpaceToken= }\\ 

\def\!thHeight{height}
\def\!thDepth{depth}
\def\!thWidth{width}

\def\!thToksEdef#1=#2{%
  \edef\!ttemp{#2}%
  #1\!thx{\!ttemp}%
  \ignorespaces}


\def\!thStoreErrorMsg#1#2{%
  \toks0 =\!thx{\csname #2\endcsname}%
  \edef#1{\the\toks0 }}

\def\!thReadErrorMsg#1{%
  \!thx\!thx\!thx\!thGobble\!thx\string #1}

\def\!thError#1#2{%
  \begingroup
    \newlinechar=`\^^J%
    \edef\!ttemp{#2}%
    \errhelp=\!thx{\!ttemp}%
    \!thMessage{%
      ^^J\!thReadErrorMsg\!thErrorMsgA 
      ^^J\!thReadErrorMsg\!thErrorMsgB}%
    \errmessage{#1}%
  \endgroup}

\!thStoreErrorMsg\!thErrorMsgA{%
  TABLE error; see manual for explanation.}
\!thStoreErrorMsg\!thErrorMsgB{%
  Type \space H <return> \space for immediate help.}

\def\!thGetReplacement#1#2{%
   \begingroup
     \!thMessage{#1}
     \endlinechar=-1
     \global\read16 to#2%
   \endgroup}


\def\!thLoop#1\repeat{%
  \def\!thIterate{%
    #1%
    \!thx \!thIterate
    \fi}%
  \!thIterate 
  \let\!thIterate\relax}


\def\Smash{%
  \relax
  \ifmmode
    \expandafter\mathpalette
    \expandafter\!thDoMathVCS
  \else
    \expandafter\!thDoVCS
  \fi}
                      
\def\!thDoVCS#1{%
  \setbox\z@\hbox{#1}%
  \!thFinishVCS}
                      
\def\!thDoMathVCS#1#2{%
  \setbox\z@\hbox{$\m@th#1{#2}$}%
  \!thFinishVCS}
                      
\def\!thFinishVCS{%
  \vbox to\z@{\vss\box\z@\vss}}






\def\!thSetDimen{%
  \ifnum \!tgCode=1
    \ifx \!tgValue\empty
      \!taDimenA \StrutHeightFactor\StrutUnit
      \advance \!taDimenA \StrutDepthFactor\StrutUnit
      \divide \!taDimenA 2
    \else
      \!taDimenA \!tgValue\StrutUnit
    \fi
  \else
    \!taDimenA \!tgValue
  \fi
  \!taDimenA=\!thSign\!taDimenA\relax
  %
  \ifmmode
    \expandafter\mathpalette
    \expandafter\!thDoMathRaise
  \else
    \expandafter\!thDoSimpleRaise
  \fi}
                      
\def\!thDoSimpleRaise#1{%
  \setbox\z@\hbox{\raise \!taDimenA\hbox{#1}}%
  \!thFinishRaise} 
                      
\def\!thDoMathRaise#1#2{%
  \setbox\z@\hbox{\raise \!taDimenA\hbox{$\m@th#1{#2}$}}%
  \!thFinishRaise}

\def\!thFinishRaise{%
  \ht\z@\z@ 
  \dp\z@\z@
  \box\z@}


\def\!thKernBack{%
  \kern -
  \ifnum \!tgCode=1 
    \ifx \!tgValue\empty 
      \the\KernFactor
    \else
      \!tgValue    
    \fi
    \KernUnit
  \else 
    \!tgValue      
  \fi
  \ignorespaces}%

\def\Vspace{%
  \noalign
  \bgroup
  \!tgGetValue\!thVspace}

\def\!thVspace{%
  \vskip
    \ifnum \!tgCode=1 
      \ifx \!tgValue\empty 
        \the\VspaceFactor
      \else
        \!tgValue    
      \fi
      \StrutUnit
    \else 
      \!tgValue      
    \fi
  \egroup} 



  
  


\def\BeginFormat{%
  \catcode`\|=12 
  \catcode`\"=12 
  \!taPreamble={}%
  \!taColumnNumber=0
  \skip0 =\InterColumnSpaceUnit
  \multiply\skip0 \InterColumnSpaceFactor
  \divide\skip0 2
  \!taRuleColumnTemplate=\!thx{%
    \!thx\tabskip\the\skip0 }%
  \!taLastRegularTabskip=\skip0 
  \!taOnceOnlyTabskipfalse
  \!taBeginFormattrue 
  \def\!tfRowOfWidths{}
  \ReadFormatKeys}

\def\!tfSetWidth{%
  \ifx \!tfRowOfWidths \empty  
    \ifnum \!taColumnNumber>0  
      \begingroup              
         \!taCountA=1          
         \aftergroup \edef \aftergroup \!tfRowOfWidths \aftergroup {%
           \aftergroup &\aftergroup \omit
           \!thLoop
             \ifnum \!taCountA<\!taColumnNumber
             \advance\!taCountA 1
             \aftergroup \!tfAOAO
           \repeat 
           \aftergroup }%
      \endgroup
    \fi
  \fi      
  \ifx [\!ttemp 
    \!thx\!tfSetWidthText
  \else
    \!thx\!tfSetWidthValue
  \fi}

\def\!tfAOAO{%
  &\omit&\omit}

\def\!tfSetWidthText [#1]{
  \def\!tfWidthText{#1}%
  \ReadFormatKeys}

\def\!tfSetWidthValue{%
  \!taMinimumColumnWidth = 
    \ifnum \!tgCode=1 
      \ifx\!tgValue\empty 
        \ColumnWidthFactor
      \else
        \!tgValue 
      \fi
      \ColumnWidthUnit
    \else
      \!tgValue 
    \fi
  \def\!tfWidthText{}
  \ReadFormatKeys}

\def\!tfSetTabskip{%
  \ifnum \!tgCode=1
    \skip0 =\InterColumnSpaceUnit
    \multiply\skip0 
      \ifx \!tgValue\empty
        \InterColumnSpaceFactor         
      \else
       \!tgValue                        
      \fi
  \else
    \skip0 =\!tgValue                   
  \fi
  \divide\skip0 by 2
  \ifnum\!taColumnNumber=0 
    \!thToksEdef\!taRuleColumnTemplate={%
      \the\!taRuleColumnTemplate 
      \tabskip \the\skip0 }
  \else
    \!thToksEdef\!taDataColumnTemplate={%
      \the\!taDataColumnTemplate 
      \tabskip \the\skip0 }
  \fi
  \if!taOnceOnlyTabskip
  \else
    \!taLastRegularTabskip=\skip0 
  \fi                             
  \ReadFormatKeys}

\def\!tfSetVrule{%
  \!thToksEdef\!taRuleColumnTemplate={%
    \noexpand\hfil
    \noexpand\vrule
    \noexpand\!thWidth
    \ifnum \!tgCode=1
      \ifx \!tgValue\empty
        \the\LineThicknessFactor      
      \else
        \!tgValue                     
      \fi
      \!taLTU                         
    \else
      \!tgValue                       
    \fi
    ####%
    \noexpand\hfil
    \the\!taRuleColumnTemplate}       
  \!tfAdjoinPriorColumn}
 
\def\!tfSetAlternateVrule{%
  \afterassignment\!tfSetAlternateA
  \toks0 =}                           

\def\!tfSetAlternateA{%
  \!thToksEdef\!taRuleColumnTemplate={%
    \the\toks0 \the\!taRuleColumnTemplate} 
  \!tfAdjoinPriorColumn}

\def\!tfAdjoinPriorColumn{%
  \ifnum \!taColumnNumber=0
    \!taPreamble=\!taRuleColumnTemplate 
    \ifnum \TracingFormats>0             
      \!tfShowRuleTemplate
    \fi
  \else
    \ifx\!tfRowOfWidths\empty  
    \else
      \!tfUpdateRowOfWidths
    \fi
    \!thToksEdef\!taDataColumnTemplate={%
      \the \!taLeftGlue
      \the \!taDataColumnTemplate
      \the \!taRightGlue}
    \ifnum \TracingFormats>0
      \!tfShowTemplates
    \fi
    \!thToksEdef\!taPreamble={%
      \the\!taPreamble
      &
      \the\!taDataColumnTemplate
      &
      \the\!taRuleColumnTemplate}
  \fi
%
  \advance \!taColumnNumber 1
  \if!taOnceOnlyTabskip              
    \!thToksEdef\!taDataColumnTemplate={%
       ####\tabskip \the\!taLastRegularTabskip}
  \else
    \!taDataColumnTemplate{##}%
  \fi
  \!taRuleColumnTemplate{}
  \!taLeftGlue{\hfil}
  \!taRightGlue{\hfil}%
  \!taMinimumColumnWidth=0pt
  \def\!tfWidthText{}%
  \!taOnceOnlyTabskipfalse    
  \ReadFormatKeys}

\def\!tfUpdateRowOfWidths{%
  \ifx \!tfWidthText\empty
  \else 
    \!tfComputeMinColWidth
  \fi
  \edef\!tfRowOfWidths{%
    \!tfRowOfWidths
    &%
    \omit                                  
    \ifdim \!taMinimumColumnWidth>0pt
      \hskip \the\!taMinimumColumnWidth
    \fi
    &
    \omit}}                                

\def\!tfComputeMinColWidth{%
  \setbox0 =\vbox{%
    \ialign{
       \span\the\!taDataColumnTemplate\cr
       \!tfWidthText\cr}}%
  \!taMinimumColumnWidth=\wd0 }

\def\!tfShowRuleTemplate{%
  \!thMessage{}
  \!thMessage{TABLE FORMAT}
  \!thMessage{Column: Template}
  \!thMessage{%
    \space *c: ##\tabskip \the\LeftTabskip}
  \!taOldRuleColumnTemplate=\!taRuleColumnTemplate}

\def\!tfShowTemplates{%
  \!thMessage{%
    \space \space r: \the\!taOldRuleColumnTemplate}
  \!taOldRuleColumnTemplate=\!taRuleColumnTemplate
  \!thMessage{%
    \ifnum \!taColumnNumber<10
      \space
    \fi
    \the\!taColumnNumber c: \the\!taDataColumnTemplate}
  \ifdim\!taMinimumColumnWidth>0pt
    \!thMessage{%
      \space \space w: \the\!taMinimumColumnWidth}
  \fi}

\def\!tfFinishFormat{%
  \ifnum \TracingFormats>0
    \!thMessage{%
      \space \space r: \the\!taOldRuleColumnTemplate
        \tabskip \the\RightTabskip}%
    \!thMessage{%
      \space *c: ##\tabskip 0pt}
  \fi
  \ifnum \!taColumnNumber<2
    \!thError{%
      \ifnum \!taColumnNumber=0
        No
      \else
        Only 1
      \fi
      "|"}%
      {\!thReadErrorMsg\!tfTooFewBarsA
       ^^J\!thReadErrorMsg\!tfTooFewBarsB
       ^^J\!thReadErrorMsg\!tkFixIt}%
  \fi
  \!thToksEdef\!taPreamble={%
    ####\tabskip\LeftTabskip 
    &
    \the\!taPreamble \tabskip\RightTabskip
    &
    ####\tabskip 0pt \cr}
  \ifnum \TracingFormats>1
    \!thMessage{Preamble=\the\!taPreamble}
  \fi
  \ifnum \TracingFormats>2
    \!thMessage{Row Of Widths="\!tfRowOfWidths"}
  \fi
  \!taBeginFormatfalse 
  \catcode`\|=13
  \catcode`\"=13
  \!ttDoHalign}

\!thStoreErrorMsg\!tfTooFewBarsA{%
  There must be at least 2 "|"'s (and/or "\string \|"'s)}
\!thStoreErrorMsg\!tfTooFewBarsB{%
  between \string\BeginFormat\space and \string\EndFormat\space (or ".").}

\def\ReFormat[{%
  \omit
  \!taDataColumnTemplate{##}%
  \!taLeftGlue{}%
  \!taRightGlue{}%
  \catcode`\|=12  
  \catcode`\"=12  
  \ReadFormatKeys}

\def\!tfEndReFormat{%
  \ifnum \TracingFormats>0
    \!thMessage{ReF: 
       \the\!taLeftGlue
       \hbox{\the\!taDataColumnTemplate}
       \the\!taRightGlue}
  \fi
  \catcode`\|=13
  \catcode`\"=13
  \!tfReFormat}

\def\!tfReFormat#1{%
  \the \!taLeftGlue
  \vbox{%
    \ialign{%
      \span\the\!taDataColumnTemplate\cr
       #1\cr}}%
  \the \!taRightGlue}







\def\!tgGetValue#1{%
  \def\!tgReturn{#1}
  \futurelet\!ttemp\!tgCheckForParen}

\def\!tgCheckForParen{%
  \ifx\!ttemp (%
    \!thx \!tgDoParen
  \else
    \!thx \!tgCheckForSpace
  \fi}

\def\!tgDoParen(#1){%
  \def\!tgCode{2}%
  \def\!tgValue{#1}
  \!tgReturn}

\def\!tgCheckForSpace{%
  \def\!tgCode{1}%
  \def\!tgValue{}
  \ifx\!ttemp\!thSpaceToken
    \!thx \!tgReturn        
  \else
    \!thx \!tgCheckForDigit         
  \fi}

\def\!tgCheckForDigit{%
  \!taDigitfalse
  \ifx 0\!ttemp
    \!taDigittrue
  \else
    \ifx 1\!ttemp
      \!taDigittrue
    \else
      \ifx 2\!ttemp
        \!taDigittrue
      \else
        \ifx 3\!ttemp
          \!taDigittrue
        \else
          \ifx 4\!ttemp
            \!taDigittrue
          \else
            \ifx 5\!ttemp
              \!taDigittrue
            \else
              \ifx 6\!ttemp
                \!taDigittrue
              \else
                \ifx 7\!ttemp
                  \!taDigittrue
                \else
                  \ifx 8\!ttemp
                    \!taDigittrue
                  \else
                    \ifx 9\!ttemp
                      \!taDigittrue
                    \fi
                  \fi
                \fi
              \fi
            \fi
          \fi
        \fi
      \fi
    \fi
  \fi
  \if!taDigit
    \!thx \!tgGetNumber
  \else
    \!thx \!tgReturn 
  \fi}

\def\!tgGetNumber{%
  \afterassignment\!tgGetNumberA
  \!taCountA=}
\def\!tgGetNumberA{%
  \edef\!tgValue{\the\!taCountA}%
  \!tgReturn}


\def\!tgSetUpParBox{%
  \edef\!ttemp{%
    \noexpand \ReadFormatKeys
    b{\noexpand \BeginTableParBox{%
      \ifnum \!tgCode=1 
        \ifx \!tgValue\empty 
          \the\ColumnWidthFactor
        \else
          \!tgValue    
        \fi
        \!taCWU        
      \else 
        \!tgValue      
      \fi}}}%
  \!ttemp
  a{\EndTableParBox}}

\def\!tgInsertKern{%
  \edef\!ttemp{%
    \kern
    \ifnum \!tgCode=1 
      \ifx \!tgValue\empty 
        \the\KernFactor
      \else
        \!tgValue    
      \fi
      \!taKU         
    \else 
      \!tgValue      
    \fi}%
  \edef\!ttemp{%
    \noexpand\ReadFormatKeys
    \ifh@            
      b{\!ttemp}
    \fi
    \ifv@            
      a{\!ttemp}
    \fi}%
  \!ttemp}




\def\NewFormatKey#1{%
  \!thx\def\!thx\!ttempa\!thx{\string #1}%
  \!thx\def\!thx\!ttempb\!thx{\csname !tk<\!ttempa>\endcsname}%
  \ifnum \TracingKeys>0
    \!tkReportNewKey
  \fi
  \!thx\ifx \!ttempb \relax
    \!thx\!tkDefineKey
  \else 
    \!thx\!tkRejectKey
  \fi}

\def\!tkReportNewKey{%
  \!taToksA\!thx{\!ttempa}%
  \!thMessage{NEW KEY: "\the\!taToksA"}}

\def\!tkDefineKey{%
  \!thx\def\!ttempb}%

\def\!tkRejectKey{%
    \!taToksA\!thx{\!ttempa}%
    \!thError{Key letter "\the\!taToksA" already used}
      {\!thReadErrorMsg\!tkFixIt}
    \def\!tkGarbage}%

\!thStoreErrorMsg\!tkFixIt{%
  You'd better type \space 'E' \space and fix your file.}


\def\ReadFormatKeys#1{%
  \!thx\def\!thx\!ttempa\!thx{\string #1}%
  \!thx\def\!thx\!ttempb\!thx{\csname !tk<\!ttempa>\endcsname}%
  \ifnum \TracingKeys>1
    \!tkReportKey
  \fi
  \!thx\ifx \!ttempb\relax 
    \!thx\!tkReplaceKey
  \else
    \!thx\!ttempb
  \fi}

\def\!tkReportKey{%
  \!taToksA\!thx{\!ttempa}%
  \!thMessage{KEY: "\the\!taToksA"}}

\def\!tkReplaceKey{%
  \!taToksA\!thx{\!ttempa}%
  \!thError {Undefined format key "\the\!taToksA"}
    {\!thReadErrorMsg\!tkUndefined ^^J\!thReadErrorMsg\!tkBadKey}
  \!tkReplaceKeyA}

\def\!tkReplaceKeyA{%
  \!thGetReplacement{\!thReadErrorMsg\!tkReplace}\!tkReplacement
  \!thx\ReadFormatKeys\!tkReplacement}

\!thStoreErrorMsg\!tkUndefined{%
  The format key in " "'s on the next to top line is undefined.}
\!thStoreErrorMsg\!tkBadKey{%
  Type \space E \space to quit now, or
  \space<CR> \space and respond to next prompt.}
\!thStoreErrorMsg\!tkReplace{%
  Type \space<replacement key><CR> \space,
   or simply \space<CR> \space to skip offending key:}


\NewFormatKey b#1{%
  \!thx\!tkJoin\!thx{\the\!taDataColumnTemplate}{#1}%
  \ReadFormatKeys}

\def\!tkJoin#1#2{%
  \!taDataColumnTemplate{#2#1}}%

\NewFormatKey a#1{%
  \!taDataColumnTemplate\!thx{\the\!taDataColumnTemplate #1}%
  \ReadFormatKeys}

\NewFormatKey \{{%
  \!taDataColumnTemplate=\!thx{\!thx{\the\!taDataColumnTemplate}}%
  \ReadFormatKeys}

\NewFormatKey *#1#2{%
  \!taCountA=#1\relax
  \!taToksA={}%
  \!thLoop 
    \ifnum \!taCountA > 0
    \!taToksA\!thx{\the\!taToksA #2}%
    \advance\!taCountA -1
  \repeat 
  \!thx\ReadFormatKeys\the\!taToksA}


\NewFormatKey \LeftGlue#1{%
  \!taLeftGlue{#1}%
  \ReadFormatKeys}

\NewFormatKey \RightGlue#1{%
  \!taRightGlue{#1}%
  \ReadFormatKeys}

\NewFormatKey c{%
  \ReadFormatKeys 
  \LeftGlue\hfil
  \RightGlue\hfil}

\NewFormatKey l{%
  \ReadFormatKeys 
  \LeftGlue{}   
  \RightGlue\hfil}

\NewFormatKey r{%
  \ReadFormatKeys 
  \LeftGlue\hfil
  \RightGlue{}}

\NewFormatKey k{%
  \h@true
  \v@true
  \!tgGetValue{\!tgInsertKern}}

\NewFormatKey i{%
  \h@true
  \v@false
  \!tgGetValue{\!tgInsertKern}}
  
\NewFormatKey j{%
  \h@false
  \v@true
  \!tgGetValue{\!tgInsertKern}}


\NewFormatKey n{%
  \def\!tnStyle{}%
   \futurelet\!tnext\!tnTestForBracket}

\NewFormatKey N{%
  \def\!tnStyle{$}%
   \futurelet\!tnext\!tnTestForBracket}


\NewFormatKey m{%
  \ReadFormatKeys b$ a$}

\NewFormatKey M{%
  \ReadFormatKeys \{ b{$\displaystyle} a$}

\NewFormatKey \m{%
  \ReadFormatKeys l b{{}} m}

\NewFormatKey \M{%
  \ReadFormatKeys l b{{}} M}

\NewFormatKey f#1{%
  \ReadFormatKeys b{#1}}

\NewFormatKey B{%
  \ReadFormatKeys f\bf}

\NewFormatKey I{%
  \ReadFormatKeys f\it}

\NewFormatKey S{%
  \ReadFormatKeys f\sl}

\NewFormatKey R{%
  \ReadFormatKeys f\rm}

\NewFormatKey T{%
  \ReadFormatKeys f\tt}

\NewFormatKey p{%
  \!tgGetValue{\!tgSetUpParBox}}


\NewFormatKey w{%
  \!tkTestForBeginFormat w{\!tgGetValue{\!tfSetWidth}}}


\NewFormatKey s{%
  \!taOnceOnlyTabskipfalse    
  \!tkTestForBeginFormat t{\!tgGetValue{\!tfSetTabskip}}}

\NewFormatKey o{%
  \!taOnceOnlyTabskiptrue
  \!tkTestForBeginFormat o{\!tgGetValue{\!tfSetTabskip}}}


\NewFormatKey |{%
  \!tkTestForBeginFormat |{\!tgGetValue{\!tfSetVrule}}}

\NewFormatKey \|{%
  \!tkTestForBeginFormat \|{\!tfSetAlternateVrule}}


\NewFormatKey .{%
  \!tkTestForBeginFormat.{\!tfFinishFormat}} 

\NewFormatKey \EndFormat{%
  \!tkTestForBeginFormat\EndFormat{\!tfFinishFormat}} 

\NewFormatKey ]{%
  \!tkTestForReFormat ] \!tfEndReFormat}


\def\!tkTestForBeginFormat#1#2{%
  \if!taBeginFormat  
    \def\!ttemp{#2}%
    \!thx \!ttemp    
  \else
    \toks0={#1}%
    \toks2=\!thx{\string\ReFormat}%
    \!thx \!tkImproperUse
  \fi}   

\def\!tkTestForReFormat#1#2{%
  \if!taBeginFormat  
    \toks0={#1}%
    \toks2=\!thx{\string\BeginFormat}%
    \!thx \!tkImproperUse
  \else
    \def\!ttemp{#2}%
    \!thx \!ttemp    
  \fi}   

\def\!tkImproperUse{%
  \!thError{\!thReadErrorMsg\!tkBadUseA "\the\toks0 "}%
    {\!thReadErrorMsg\!tkBadUseB \the\toks2 \space command.
    ^^J\!thReadErrorMsg\!tkBadKey}%
  \!tkReplaceKeyA}
 
\!thStoreErrorMsg\!tkBadUseA{Improper use of key }  
\!thStoreErrorMsg\!tkBadUseB{%
  The key mentioned above can't be used in a }




\def\!tnTestForBracket{%
  \ifx [\!tnext
    \!thx\!tnGetArgument
  \else
    \!thx\!tnGetCode
  \fi}

\def\!tnGetCode#1 {
  \!tnConvertCode #1..!}

\def\!tnConvertCode #1.#2.#3!{%
  \begingroup
    \aftergroup\edef \aftergroup\!ttemp \aftergroup{%
      \aftergroup[%
      \!taCountA #1
      \!thLoop
        \ifnum \!taCountA>0
        \advance\!taCountA -1
        \aftergroup0
      \repeat
      \def\!ttemp{#3}%
      \ifx\!ttemp \empty
      \else
        \aftergroup.
        \!taCountA #2
        \!thLoop 
          \ifnum \!taCountA>0
          \advance\!taCountA -1
          \aftergroup0
        \repeat
      \fi 
      \aftergroup]\aftergroup}%
    \endgroup\relax
    \!thx\!tnGetArgument\!ttemp}
  
\def\!tnGetArgument[#1]{%
  \!tnMakeNumericTemplate\!tnStyle#1..!}

\def\!tnMakeNumericTemplate#1#2.#3.#4!{
  \def\!ttemp{#4}%
  \ifx\!ttemp\empty
    \!taDimenC=0pt
  \else
    \setbox0=\hbox{\m@th #1.#3#1}%
    \!taDimenC=\wd0
  \fi
  \setbox0 =\hbox{\m@th #1#2#1}%
  \!thToksEdef\!taDataColumnTemplate={%
    \noexpand\!tnSetNumericItem
    {\the\wd0 }%
    {\the\!taDimenC}%
    {#1}%
    \the\!taDataColumnTemplate}  
  \ReadFormatKeys}

\def\!tnSetNumericItem #1#2#3#4 {
  \!tnSetNumericItemA {#1}{#2}{#3}#4..!}

\def\!tnSetNumericItemA #1#2#3#4.#5.#6!{%
  \def\!ttemp{#6}%
  \hbox to #1{\hss \m@th #3#4#3}%
  \hbox to #2{%
    \ifx\!ttemp\empty
    \else
       \m@th #3.#5#3%
    \fi
    \hss}}




\def\MakeStrut#1#2{%
  \vrule width0pt height #1 depth #2}

\def\StandardTableStrut{%
  \MakeStrut{\StrutHeightFactor\StrutUnit}
    {\StrutDepthFactor\StrutUnit}}

\def\AugmentedTableStrut#1#2{%
  \dimen@=\StrutHeightFactor\StrutUnit
  \advance\dimen@ #1\StrutUnit
  \dimen@ii=\StrutDepthFactor\StrutUnit
  \advance\dimen@ii #2\StrutUnit
  \MakeStrut{\dimen@}{\dimen@ii}}

\def\Enlarge#1#2{
  \!taDimenA=#1\relax
  \!taDimenB=#2\relax
  \let\!TsSpaceFactor=\empty
  \ifmmode
    \!thx \mathpalette
    \!thx \!TsEnlargeMath
  \else
    \!thx \!TsEnlargeOther
  \fi}

\def\!TsEnlargeOther#1{%
  \ifhmode
    \setbox\z@=\hbox{#1%
      \xdef\!TsSpaceFactor{\spacefactor=\the\spacefactor}}%
  \else
    \setbox\z@=\hbox{#1}%
  \fi
  \!TsFinishEnlarge}
    
\def\!TsEnlargeMath#1#2{%
  \setbox\z@=\hbox{$\m@th#1{#2}$}%
  \!TsFinishEnlarge}

\def\!TsFinishEnlarge{%
  \dimen@=\ht\z@
  \advance \dimen@ \!taDimenA
  \ht\z@=\dimen@
  \dimen@=\dp\z@
  \advance \dimen@ \!taDimenB
  \dp\z@=\dimen@
  \box\z@ \!TsSpaceFactor{}}


\def\OpenUp#1#2{%
  \advance \StrutHeightFactor #1\relax
  \advance \StrutDepthFactor #2\relax}




\def\BeginTable{%
  \futurelet\!tnext\!ttBeginTable}

\def\!ttBeginTable{%
  \ifx [\!tnext
    \def\!tnext{\!ttBeginTableA}%
  \else 
    \def\!tnext{\!ttBeginTableA[c]}%
  \fi
  \!tnext}

\def\!ttBeginTableA[#1]{%
  \if #1u
    \ifmmode                 
      \def\!ttEndTable{
        \relax}
    \else                   
      \bgroup
      \def\!ttEndTable{%
        \egroup}%
    \fi
  \else
    \hbox\bgroup $
    \def\!ttEndTable{%
      \egroup 
      $
      \egroup}
    \if #1t%
      \vtop
    \else
      \if #1b%
        \vbox
      \else
        \vcenter 
      \fi
    \fi
    \bgroup      
  \fi
  \advance\!taRecursionLevel 1 
  \let\!ttRightGlue=\relax  
  \everycr={}
  \ifnum \!taRecursionLevel=1
    \!ttInitializeTable
  \fi}

\bgroup
  \catcode`\|=13
  \catcode`\"=13
  \catcode`\~=13
  \gdef\!ttInitializeTable{%
    \let\!ttTie=~ 
    \let\!ttDH=\- 
    \catcode`\|=\active
    \catcode`\"=\active
    \catcode`\~=\active
    \def |{\unskip\!ttRightGlue&&}
    \def\|{\unskip\!ttRightGlue&\omit\!ttAlternateVrule}%
    \def"{\unskip\!ttRightGlue&\omit&}
    \def~{\kern .5em}
    \def\\{\!ttEndOfRow}%
    \def\-{\!ttShortHrule}%
    \def\={\!ttLongHrule}%
    \def\_{\!ttFullHrule}%
    \def\Left##1{##1\hfill\null}
    \def\Center##1{\hfill ##1\hfill\null}
    \def\Right##1{\hfill##1}%
    \the\EveryTable}
\egroup

\let\!ttRightGlue=\relax  

\def\!ttDoHalign{%
  \baselineskip=0pt \lineskiplimit=0pt \lineskip=0pt %
  \tabskip=0pt
  \halign \the\!taTableSpread \bgroup
   \span\the\!taPreamble
   \ifx \!tfRowOfWidths \empty
   \else 
     \!tfRowOfWidths \cr %
   \fi}

\def\EndTable{%
  \egroup 
  \!ttEndTable}


\def\!ttEndOfRow{%
  \futurelet\!tnext\!ttTestForBlank}

\def\!ttTestForBlank{%
  \ifx \!tnext\!thSpaceToken  
    \!thx\!ttDoStandard
  \else
    \!thx\!ttTestForZero
  \fi}
  
\def\!ttTestForZero{%
  \ifx 0\!tnext
    \!thx \!ttDoZero
  \else
    \!thx \!ttTestForPlus
  \fi}

\def\!ttTestForPlus{%
  \ifx +\!tnext
    \!thx \!ttDoPlus
  \else
    \!thx \!ttDoStandard
  \fi}

\def\!ttDoZero#1{
  \cr} 

\def\!ttDoPlus#1#2#3{
  \AugmentedTableStrut{#2}{#3}%
  \cr} 

\def\!ttDoStandard{%
  \StandardTableStrut
  \cr}


 



\def\!ttAlternateVrule{%
  \!tgGetValue{\!ttAVTestForCode}}  

\def\!ttAVTestForCode{%
  \ifnum \!tgCode=2              
    \!thx\!ttInsertVrule         
  \else
    \!thx\!ttAVTestForEmpty
  \fi}

\def\!ttAVTestForEmpty{%
  \ifx \!tgValue\empty           
    \!thx\!ttAVTestForBlank
  \else
    \!thx\!ttInsertVrule         
  \fi}

\def\!ttAVTestForBlank{%
  \ifx \!ttemp\!thSpaceToken     
    \!thx\!ttInsertVrule
  \else
    \!thx\!ttAVTestForStar 
  \fi}

\def\!ttAVTestForStar{%
  \ifx *\!ttemp                  
    \!thx\!ttInsertDefaultPR     
  \else
    \!thx\!ttGetPseudoVrule       
  \fi}

\def\!ttInsertVrule{%
  \hfil 
  \vrule \!thWidth
    \ifnum \!tgCode=1
      \ifx \!tgValue\empty 
        \LineThicknessFactor
      \else
        \!tgValue
      \fi
      \LineThicknessUnit
    \else
      \!tgValue
    \fi
  \hfil
  &}

\def\!ttInsertDefaultPR*{%
  \PseudoVrule    
  &}

\def\!ttGetPseudoVrule#1{%
  \toks0={#1}%
  #1&}

\def\PseudoVrule{}


\def\!ttuse#1{%
  \ifnum #1>\@ne 
    \omit 
    \mscount=#1 
    \advance\mscount by \m@ne
    \advance\mscount by \mscount
    \!thLoop 
      \ifnum\mscount>\@ne 
      \sp@n 
    \repeat 
    \span 
  \fi}

\def\!ttUse#1[{%
  \!ttuse{#1}%
  \ReFormat[}


\def\!ttFullHrule{%
  \noalign
  \bgroup
  \!tgGetValue{\!ttFullHruleA}}

\def\!ttFullHruleA{%
  \!ttGetHalfRuleThickness 
  \hrule \!thHeight \dimen0 \!thDepth \dimen0
  \penalty0 
  \egroup} 

\def\!ttShortHrule{%
  \omit
  \!tgGetValue{\!ttShortHruleA}}

\def\!ttShortHruleA{%
  \!ttGetHalfRuleThickness 
  \leaders \hrule \!thHeight \dimen0 \!thDepth \dimen0 \hfill
  \null    
  \ignorespaces} 

\def\!ttLongHrule{%
  \omit\span\omit\span \!ttShortHrule}

\def\!ttGetHalfRuleThickness{%
  \dimen0 =
    \ifnum \!tgCode=1
      \ifx \!tgValue\empty
        \LineThicknessFactor
      \else
        \!tgValue    
      \fi
      \LineThicknessUnit
    \else
      \!tgValue      
    \fi
  \divide\dimen0 2 }


\def\SetTableToWidth#1{%
  \!taTableSpread={to #1}}

\def\WidenTableBy#1{%
  \ifdim #1=0pt
    \!taTableSpread={}%
  \else
    \!taTableSpread={spread #1}%
  \fi}

%


\def\JustLeft{%
  \omit \let\!ttRightGlue=\hfill}
\def\JustCenter{%
  \omit \hfill\null \let\!ttRightGlue=\hfill}

\let\\=\!tacr
\catcode`\!=12
\catcode`\@=12

\overfullrule=0pt
\headline={\ifnum\pageno=1\firstheadline\else
\ifodd\pageno\rightheadline \else\leftheadline\fi\fi}
\def\firstheadline{\hfil}
\def\rightheadline{\hfil}
\def\leftheadline{\hfil}
	\footline={\ifnum\pageno=1\firstfootline\else\otherfootline\fi}
\def\firstfootline{\rm\hss\folio\hss}
\def\otherfootline{\hfil}

 1
 1
 1

\font\tenbf=cmbx10
\font\tenrm=cmr10
\font\tenit=cmti10
\font\ninebf=cmbx9
\font\ninerm=cmr9
\font\nineit=cmti9
\font\eightbf=cmbx8
\font\eightrm=cmr8
\font\eightit=cmti8

\parindent=1.2pc
\magnification=\magstep1
\hsize=6.0truein
\vsize=8.6truein
\nopagenumbers

\centerline{\tenbf CRITICAL REVIEW OF THE ELECTROWEAK PRECISION TESTS\footnote"$^*$"{\eightrm\baselineskip=10pt
Presented by K.K. at the International Symposium on Heavy Flavor and
Electroweak Theory, Beijing, China, 17 - 19 August 1995, and supported
in part by the US DOE(KK) and also KOSEF through SNU CTP(SKK). BROWN-HET-1033,
BROWN-TA-532 and hep-ph/9602261.}}
\baselineskip=13pt

\vglue0.8cm
\centerline{\eightrm KYUNGSIK KANG}
\baselineskip=12pt
\centerline{\eightit Physics Department, Brown University,}
\baselineskip=10pt
\centerline{\eightit Providence, Rhode Island 02912, U.S.A.}
\centerline{\eightrm E-mail: kang\@het.brown.edu}
\vglue0.2cm
\centerline{\eightrm and}
\vglue0.2cm
\centerline{\eightrm SIN KYU KANG}
\baselineskip=12pt
\centerline{\eightit Center for Theoretical Physics, Seoul National University, Seoul 151-742, Korea}
\centerline{\eightrm E-mail: skkang\@phyb.snu.ac.kr}
\vglue0.2cm

\vglue0.6cm
\centerline{\eightrm ABSTRACT}
\vglue0.2cm
{\rightskip=3pc
 \leftskip=3pc
 \eightrm\baselineskip=10pt\noindent
There have been a great deal of works on the precision test of the standard
model (SM) because of the incredibly precise data obtained at the LEP and 
the new measurements of $M_W$ and $m_t$ at the Fermilab Tevatron
as well as the recent theoretical 
progress in the higher order radiative corrections. We will discuss some
of the hidden inputs and theoretical uncertainties involved in making the
predictions of the observables in the SM. From the minimal $\chi^2$-fit to the experimental Z-decay parameters (with the aid of a modified ZFITTER program) in a scheme where $M_Z$, $G_{\mu}$ and $\alpha (M_Z)$ are 
taken as inputs, we can predict $M_W$ for given values of $m_H$ and $m_t$. 
The current world average value of $M_W$  definitely favor nonvanishing electroweak radiative corrections and is consistent with a heavy $m_t$ as measured by the recent CDF report but with a heavy Higgs scalar of about 
500 GeV within the context  of the minimal SM. 
The sensitivity of and the errors in the best fit solutions 
due to the uncertainties in the gluonic coupling $\alpha_s(M_Z)$ and 
$\alpha (M_Z)$ are examined carefully and any trace of new physics beyond the SM implied by the data, in a particular $R_b$ and $R_c$, is also touched upon. In addition we discuss how the future precision measurements of $M_W$ and $Z$ decay parameters can determine the Higgs boson mass and distinguish the SM with radiative corrections from the minimal supersymmetric standard model (MSSM) and other extended model. 

\vglue0.6cm}

\tenrm\baselineskip=13pt
\leftline{\tenbf 1. Introduction}
The discovery~$^1$ of the t-quark presents yet another challenge for the precision tests of the electroweak standard model of the leptons and colored quarks. Namely, one can now examine critically the uncertainties in the predicted $M_W$ mass and the $Z-$decay parameters.
Within the framework of the standard model in which $G_{\mu}, \alpha$ and
$M_Z$ are taken as input, one can predict $M_W$ from the mass relation with the radiative corrections as well as the $Z-$decay parameters. 
Starting with the given masses of the quarks, leptons, gauge bosons and Higgs
scalar, as well as given gluonic coupling $\alpha_s(M_Z)$ and gauge coupling
 $\alpha(M_Z)$, the radiative correction
$\Delta r$ can be calculated by including up to the dominant two-loop and
QCD-electroweak mixed terms, which in turn can be used in the $W-$mass relation to determine $M_W$ in a self-consistent manner. The most important inputs in terms of causing significant uncertainties to the physical observables such as $M_W$ and $Z-$decay parameters in this procedure are $m_t$, $\alpha (M_Z)$ and
$\alpha _s (M_Z)$. We will discuss some of the hidden issues of the electroweak precision tests and theoretical uncertainties of the predicted observables as well as some issues concerning the test of the new physics beyond the SM. 
We begin with listing some of the recent experimental and theoretical progress on the electroweak parameters: 
\item [1.] The most recent values~$^1$ of the t-quark mass
are $m_t = 176 {\pm} 8 {\pm} 10$ GeV (CDF) and 
$m_t = 199 ^{+19}_{-21} {\pm} 22$ GeV (D0).
\item [2.] The most recent CDF result~{$^2$ of the W-boson mass is
$M_W = 80.41 {\pm} 0.18 $ GeV which together with the D0 and other results makes the '95 world average~$^3$ $M_W = 80.26 {\pm} 0.16$ GeV.
\item [3.] The improved LEP results~$^{3,4}$ on Z-decay parameters and $M_Z$.

The experimental precision of the electroweak data has gotten improved steadily during the past several years. As a representative example, the mass of the Z-boson is now $M_Z = 91.1884 {\pm} 0.0022$ GeV compared to the '93 value $91.187 {\pm} 0.007$ GeV and the '94 value $91.1888 {\pm} 0.0044$ GeV. The accuracy of the total and some of the partial decay widths of the Z-boson is at the level of a few factor of $10^{-3}$ so that not only the quantum electroweak corrections can now be probed but also the uncertainty in the QED running coupling constant at the Z-boson mass scale is an appreciable source of the errors in the precision test of the standard and minimally extended models. 

In addition, some of the theoretical progress on the higher order corrections are:
\item [4.] the dominant two-loop corrections~$^5$ of the order $O ({\alpha ^2} m_t ^2)$ to $\Delta \rho = 1 - 1 / {\rho} $, and
\item [5.] the QCD corrections to the leading electroweak one-loop term~$^{5,6}$ which are of the orders $O (\alpha \alpha_s m_t ^2)$ and 
$O (\alpha \alpha_s ^2 m_t ^2)$. This affects the $G_{\mu} m_t ^2$ term in $\Delta \rho$. We note however that the higher order QCD effects of $O (\alpha \alpha_s ^2 m_t ^2)$ are yet to be settled unambiguously~$^{6,7}$.

The W-boson mass relaton will be affected by all these corrections through $\Delta r$. The total and partial Z-decay widths receive all these corrections through $\Delta r$ as well as the higher order QCD corrections through the QCD factor $R_{QCD}$ , for which we use the results~$^{8,9}$ of up to the three-loop order calculations with mass dependent coefficients. 

  We would like to present the results of the new fit to the {\it updated}
1995 data with the aid of the appropriately modified  ZFITTER program~$^{10}$ 
to incorporate these new experimental and theoretical developments.~$^{11}$
We examine the uncertainties in the best fit solutions of the Z-decay parameters and the predicted $M_W$ due to the current errors in 
$\alpha_s (M_Z)$ and $\alpha (M_Z)$ as well as in $m_t$.

  In the analysis we determine $M_W$ self-consistently from the W-mass relation that includes the electroweak radiative corrections (EWRC) for the value of $m_t$ covering the experimental range and fit the LEP data. We will see how stable the predicted $M_W$ is regardless of the exact value of $m_H$ in the interesting range of $60-1000$ GeV.
The sensitivity of the EWRC to the exact value of $M_W$
in the standard model has been studied based on the $W$-mass formula.~$^{12}$
Also we will see how sensitive the precision tests and the $m_t-m_H$ correlation are to the requirement of self-consistency in the W-mass relation with the needed EWRC as well as to the errors in 
$\alpha_s(M_Z)$ and $\alpha (M_Z)$. We examine critically how consistent the best-fit solutions, i.e., the minimal $\chi ^2$ solutions, are with the CDF $m_t$ for the best determined values of $\alpha _s (M_Z)$ and 
$\alpha ^{-1} (M_Z)$ and what range of the Higgs boson mass and $M_W$ are implied by the best fit solutions as well as by the uncertainties due to the errors in the strong and QED coupling constants. 
In addition, we examine the validity of
the QED Born approximation (QBA)~$^{13}$ in which $\alpha (M_Z)$
is used instead of $\alpha(0)$ in the tree approximation
 along with the corresponding redefinition of the weak
mixing angle $\sin ^2\theta $ instead of $\sin ^2\theta_W $. 

The electroweak parameters relevant to the precision tests within the framework of the SM are introduced in the next Section and the numerical results are presented in Section 3. Section 4 contains conclusions and remarks on the precision tests of the SM as well as on the possible indication of new physics beyond the SM that may be implied by the current precision electroweak data.

\vglue0.4cm
\leftline{\tenbf 2. Electroweak Parameters}
It is well known that the charge renormalization in the conventional QED
fixes the counter term by
the renormalized vacuum polarization $\hat {\Pi}^{\gamma }(0)$ and
one can evaluate
$\hat {\Pi}^{\gamma}(q^2)=\hat {\Sigma} ^{\gamma \gamma }(q^2)/q^2$
from the photon self energy $\hat{\Sigma}^{\gamma \gamma }(q^2)$, for example,
by the dimensional regularization method. This gives at $q^2=M_Z^2$
the total fermionic contribution of $m_f \leq M_Z $ to the real part
$\Delta \alpha = - Re\hat {\Pi}^{\gamma}(M_Z^2) =  0.05944(65) $,
which includes both the lepton and quark parts.~$^{14}$
Here, the quark contribution to $Re\hat {\Pi }^{\gamma}(q^2)$ is
the hadronic one which can be directly evaluated  by dispersion
integral over the measured cross section of $e^+e^- \rightarrow hadrons$.
Then, we get from
$\alpha (M_Z) =  \alpha / (1 - \Delta \alpha)$ that $\alpha^{-1} (M_Z) = 128.89(9)$ in the on-shell scheme
if the hyperfine structure constant $\alpha = e^2/4\pi = 1/137.0359895(61)$
is used, which is  The value in agreement with the two most recent calculations.~$^{14,15}$ 
The error in $\alpha^{-1} (M_Z)$ is essentially due to the uncertainty in hadronic contribution.
The electroweak parameters are evaluated
numerically with the hyperfine structure constant $\alpha$,
the four-fermion coupling constant of $\mu$-decay,
$G_{\mu} = 1.16639(2)\times 10^{-5} ~ ~ ~{GeV}^{-2}$, and $Z$-mass
$M_Z = 91.1884(22)$ in the 1995 data fit.
Numerical estimate of the full EWRC requires
the mass values of the leptons, quarks, and Higgs scalar besides these
quantities. The minimal $\chi^2$-fit to the LEP data will at best give  
$m_t - m_H$ correlation. The best-fit solutions are chosen out of the solution set for $(m_t, m_H)$ and $M_W$ is determined self-consistently from the $W$-mass
relation with EWRC. 

One has in the SM the on-shell
relation $\sin^2 \theta_W = 1-\frac{M_W^2}{M_Z^2}$, while the four-fermion coupling constant $G_{\mu}$ can be written as
$$G_{\mu } = \frac{\pi \alpha }{\sqrt{2}M_W^2}
\left(1-\frac{M_W^2}{M_Z^2}\right)^{-1}(1-\Delta r)^{-1} \eqno(1)$$
so that $\Delta r$, representing the radiative corrections, is given by
$$\Delta r = 1-\left(\frac{37.2802}{M_W}\right)^2
\frac{1}{1-M_W^2/M_Z^2}. \eqno(2)$$
We note that the radiative correction $\Delta r$ is very sensitive
to the value of $M_W$.
Mere change in $M_W$ by $0.44\%$ results as much as a $40\%$ change in
$\Delta r$. Theoretically, the radiative correction parameter $\Delta r$ within the SM can be written as~$^{16}$
$$1 - \Delta r = (1 - \Delta \alpha ) (1 + \cot^2 \theta_W \Delta \rho )
- \Delta r_{rem} , \eqno(3)$$
where $\Delta \rho$ contains one-loop and the leading two-loop irreducible weak and QCD corrections.
Any other corrections than $\Delta \alpha $ would represent
the genuine electroweak radiative effects.
Main contribution to $\Delta \rho $ is from the heavy t-quark
through the mass renormalizations of weak gauge bosons $W $ and $Z,$
while there is a part in $(\Delta r)_{rem}$ containing also t-quark
and Higgs scalar contributions. Note that the so-called QBA to $\Delta r$ is defined by keeping only the photon vacuum polarization contribution,
$\Delta \alpha = -Re \hat{\Pi }^{\gamma}(M_Z^2) = 0.05944$. 
We note that $\Delta \alpha $ is numerically the
dominant component of the radiative corrections, i.e., with the current world average value $M_W=80.26 $ GeV,
$\Delta \alpha $ differs by mere $29\%$ from the needed $\Delta r$ that has
to be accounted for by the weak interaction corrections.~$^{12}$ 

Starting with the given masses of the quarks, leptons, gauge bosons and Higgs
scalar, as well as given gluonic coupling $\alpha_s(M_Z)$ and gauge coupling
 $\alpha(M_Z)$,
$\Delta r$ is calculated from (3) by including up to the dominant two-loop and
QCD-electroweak mixed terms and then is used to determine $M_W$ from the 
right hand side of (2). With this new $M_W$, $\Delta r$ is calculated again
to determine another new $M_W$.
This iteration process is repeated until $\Delta r $ converges
to within $O(10^{-6})$. The final output $M_W$ from this iteration procedure 
is the self-consistent solution of (2) for $M_W$ with the starting set of
$m_t, m_H$ etc.
Upon varying $m_t$, this procedure will give the $m_t-M_W$ correlation
for all other parameters including $m_H$ fixed.
We then calculate the eleven $Z$-decay parameters, as chosen in
Table 1 in the next Section, for the parameter sets ($m_t, m_H$) that 
determine $M_W$ from (2) 
and search for the minimal $\chi^2$-fit solution to the experimental 
$Z$-decay parameters.
This procedure selects the best-fit solution. 

The $Z$-decay parameters are calculated
with the gluonic coupling constant in the range
$\alpha _s(M_Z) = 0.123(6)$
in the QCD correction factor
 $$R_{QCD} = 1+1.05\frac{\alpha_s}{\pi}+0.9(\pm 0.1)
\left(\frac{\alpha_s}{\pi}\right)^2-13.0
\left(\frac{\alpha_s}{\pi}\right)^3 \eqno(4)$$
for the light quarks(u,d,c,s)~$^8$ and in
$$R_{QCD} = 1+c_1(m_b)\frac{\alpha_s}{\pi}+c_2(m_t,m_b)
\left(\frac{\alpha_s}{\pi}\right)^2+c_3(m_t,m_b)
\left(\frac{\alpha_s}{\pi}\right)^3 , \eqno(5)$$ 
with the $m_b$ and $m_t$ mass dependent coefficients for the b quark.~$^9$
The partial width for $Z\rightarrow f\bar{f}$ is given by
$$\Gamma_f = \frac{G_{\mu}}{\sqrt{2}}\frac{M_Z^3}{24\pi}\beta R_{QED}c_fR_{QCD}(M_Z^2)\left \{[(\bar{v}^Z_f)^2+(\bar{a}^Z_f)^2] \left(1+2\frac{m_f^2}{M_Z^2}\right)-6(\bar{a}^Z_f)^2\frac{m_f^2}{M_Z^2}
\right \} \eqno(6)$$
where $\beta =\sqrt{1-4m_f^2/M_Z^2}$, $R_{QED}
=1+\frac{3}{4}\frac{\alpha}{\pi}Q_f^2$ and the color factor $c_f=3$ for
quarks and 1 for leptons.
Here the renormalized vector and axial-vector couplings are defined by
$\bar{a}_f^Z=\sqrt{\rho_f^Z}2a_f^Z = \sqrt{\rho_f^Z}2I_3^f $ and
$\bar{v}^Z_f=\bar{a}^Z_f[1-4|Q_f|\sin^2\theta_W\kappa^Z_f] $ in
terms of the familiar notations.~$^{10,16,17}$
Note that $\Delta \alpha $  is included in the
couplings through $\sin^2 \theta_W$ via (1) and (3) and all other
non-photonic loop corrections are
grouped in $\rho_f^Z$ and $\kappa_f^Z$, as in Ref.10 and 18,
including the dominant two-loop and QCD-electroweak terms.
Note that the QED loop corrections can unambiguously be separated from
the electroweak loops in the case of neutral current interactions.~$^{16,17}$
Thus the case of the QBA
can be achieved simply by setting $\rho^Z_f$ and $\kappa^Z_f$
to 1 in the vector and axial-vector couplings.

\vglue0.4cm
\leftline{\tenbf 3. Numerical Results}
The numerical results for the minimal $\chi^2$-fits to the updated 1995 
LEP data for eleven Z-decay parameters and theoretically determined 
$M_W$ are given in Table 1. They are significantly
different from the fits~$^{19}$ to the 1993 data, though similar to those~$^{11}$ of the 1994 data except for $R_b = R(\Gamma_{b\bar b} / \Gamma_{had})$ and $R_c = R(\Gamma_{c\bar c} / \Gamma_{had})$. The new values for these parameters are now  $R_b$ = 0.2219(17) and $R_c$ = 0.1543(74)
compared to the 1994 values 0.2202(20) and 0.1583(98). They are about 3.5 $\sigma$ and 2.3 $\sigma$ away from the SM predictions of the best fit solutions respectively and are the main source of much larger $\chi^2$ values in the case of the 1995 data. 
There is however a clear evidence of the full EWRC in each of the eleven $Z$-parameters. The best fit solutions to the 1995 data show
a stable output $M_W$ = 80.29(3)  GeV for $m_H$ in the range of 
$60 - 1000$ GeV, while the $\chi^2$ values favor the lower limit of $M_W$
and accordingly lower $m_t$ than the CDF value and a lighter $m_H$.
In particular 
the QBA gives inferior $\chi^2$ (=23.0/11) for the  1995 data, which is
comparable to the case ($m_t$ = 186 GeV, $m_H$ = 1000 GeV). However in the case
of QBA one gets $M_W$ = 79.96(9) GeV and $\Delta r$ = 0.0596(9) to be compared to $M_W$ = 80.29(3) GeV for $m_H$ = 60 - 1000 GeV and $\Delta r$ = 0.0390(9) -
0.0423(11), while the required  $\Delta r$ = 0.0443(102).
Also the CDF $m_t=176$ GeV is a possible output solution
with a $m_H $ about 500  GeV
among the many possible combinations of $(m_t, m_H)$. As shown in Table 1, the best-fit solutions can have errors due to the 
uncertainty in $\alpha_s(M_Z)$ : $m_t$ and $M_W$ may be shifted by as much as
$\pm 5$ GeV and $\pm 30 $ MeV respectively because of $\Delta \alpha_s=\pm 0.006$. There are additional comparable error due to the uncertainty in 
$\alpha (M_Z)$ as shown in Table 1: $\Delta \alpha^{-1}(M_Z)=\pm 0.09$ can cause another $\pm 5$ GeV
and $\pm 20 $ MeV respectively in $m_t$ and $M_W$.

Though the $\chi^2$-values tend to prefer the lower $m_t$ 
 and accordingly smaller $m_H$, there are infinitly many pairs of $(m_t, m_H)$ all of which are from a minimal $\chi^2$ and 
 statistically comparable to each other. In particular the best global fits to the updated 1995 data give $m_t = 163 - 187$ GeV for $m_H = 200 - 1000$ GeV.
Most of the $Z$-parameters are stable irrespectively to the uncertainties due
to $\Delta \alpha_s $ and $\Delta \alpha $ and in excellent agreement with 
the data except $R_b$ and $R_c$. Even with 
the mass dependent QCD factor, there is still about 3.5 $\sigma$ and 2.3  $\sigma$ deviations respectively in $R_b$ and $R_c$ from the experiments irrespectively to
the uncertainties in $\alpha_s(M_Z)$.~$^{6}$
 Most of the $\chi^2 $ contributions are from $R_b$ and to a lesser degree from $R_c$ and $A^{0,l}_{FB}$.

$M_W$ changes with $m_t$ for fixed $m_H$ from  the consistency
of the full EWRC and the central values of the world average  $M_W$ and CDF $m_t$ are consistent with a Higgs scalar mass about 1000 GeV, 
though $m_H = 100$ GeV is within 1 $\sigma$ because of large errors in the data.
Clearly a better precision measurement of $M_W$ is desired to distinguish
different $m_H$. For example, a change of $m_H$ by 200 GeV, i.e., from 500 GeV to 300 GeV
at $m_t=176$ GeV, results a change of 44 MeV in $M_W$, i.e.,
from 80.326 GeV to 80.37 GeV. This in turn will require a
 precision of 10 GeV or better in $m_t$ from the best-fit solution,
 which is consistent with the most statistical error improvement that may be 
achieved at the Fermilab Tevatron.
Present precisions in the data entail a theoretical uncertainty of about
36 MeV in $M_W$ which is about the overall error improvement expected at 
LEP-200.
\vfill
\eject
\midinsert
{\eightpoint Table~1. Numerical results including full EWRC for
eleven experimental parameters of the Z-decay and $M_W$. 
Each pair of $m_t$ and $m_H$ represents the case of the best $\chi ^2$-
fit to the 1995 LEP data for $\alpha_s(M_Z)
=0.123(6) $ and $\alpha^{-1}(M_Z)=128.89(9)$. 
The numbers in () represent the errors due to $\Delta \alpha_s(M_Z) = 
\pm 0.006$ and $\Delta \alpha^{-1}(M_Z)=\pm 0.09$ respectively.}
\vglue8pt
\noindent
$$\BeginTable
\ninepoint

\def\PseudoVrule{\hfil\vrule\hskip2pt\vrule\hfil}
\EveryTableParBox={\noindent\tolerance=10000\hbadness=10000}	
\SetTableToWidth{6.5truein}
\BeginFormat
|c|c|c|c|c|c|.
\_
\|* |Experiment | Full EW | Full EW | Full EW | Full EW\|*\\ 
\_
\|* | | | | | \|*\\
\|*$m_t$~(GeV) | $176\pm 10^{+13}_{-12}$  | $186^{(4)(4)}_{(4)(4)}$ | 
 $176^{(4)(4)}_{(4)(5)}$|
  $168^{(4)(4)}_{(4)(5)}$ | $146^{(5)(5)}_{(5)(5)}$ \|*\\
\|* | | | | | \|*\\
\|*$ m_H$~(GeV) | 60 $\leq m_H \leq 1000$ | $1000$ |
 $500$ | 300 | 60 \|*\\
\|* | | | | | \|*\\
\_
\|* | | | | | \|*\\
\|*$M_W$~(GeV) | $80.26\pm 0.16$ |
$ 80.32^{(3)(1)}_{(3)(2)}$ |$80.31^{(3)(1)}_{(3)(2)}$ | 
$80.30^{(2)(1)}_{(3)(2)}$ | $80.26^{(3)(2)}_{(2)(1)}$ \|*\\
\|* | | | | | \|*\\
\|*$ \Gamma_Z $~(MeV) | $2496.3\pm 3.2 $ |$2495.7^{(2.1)(0.5)}_{(2.2)(0.6)}$
| $2495.9^{(1.9)(0.3)}_{(2.2)(0.6)}$ |
$2495.8^{(1.9)(0.5)}_{(2.2)(0.7)}$ |
$ 2494.4^{(2.3)(0.6)}_{(2.1)(0.6)}$ \|*\\
\|* | | | | | \|*\\
\|*$ \sigma_{h}^P(nb)$ | $41.488\pm 0.12 $
| $41.449^{(33)(3)}_{(33)(3)}$ |
$41.442^{(33)(3)}_{(33)(3)}$| $41.437^{(33)(3)}_{(32)(3)}$ | 
$ 41.425^{(33)(3)}_{(32)(3)}$ \|*\\
\|* | | | | | \|*\\
\|*$R(\Gamma_{had}/\Gamma_{l\bar{l}})$ | $20.788 \pm 0.032$ | 
 $20.761^{(39)(4)}_{(40)(6)}$ |
  $20.769^{(40)(5)}_{(39)(4)}$ | $20.776^{(39)(4)}_{(40)(5)}$ |
$ 20.792^{(40)(5)}_{(39)(4)}$\|*\\
\|* | | | | | \|*\\  
\|*$ A^{0,l}_{FB}$ | $ 0.0172\pm 0.0012 $|
$ 0.01537^{(28)(12)}_{(27)(12)}$ | 
$0.01540^{(27)(7)}_{(33)(12)}$| $0.01541^{(26)(7)}_{(32)(13)}$ | 
$0.01557^{(30)(10)}_{(24)(10)}$ \|*\\
\|* | | | | | \|*\\ 
\|*$ A_{\tau} $ | $ 0.1418\pm 0.0075$ |
$0.1432^{(13)(5)}_{(13)(6)}$ |
$0.1433^{(13)(3)}_{(15)(6)}$| 
$0.1433^{(13)(4)}_{(14)(6)}$ | $0.1441^{(14)(4)}_{(11)(5)}$ \|*\\
\|* | | | | | \|*\\ 
\|*$ A_{e} $ | $ 0.1390\pm 0.0089$|
$0.1432^{(13)(5)}_{(13)(6)}$ |
$0.1433^{(13)(3)}_{(15)(6)}$| 
$0.1433^{(13)(4)}_{(14)(6)}$ | $0.1441^{(14)(4)}_{(11)(5)}$ \|*\\
\|* | | | | | \|*\\
\|*$R(\Gamma_{b\bar{b}}/\Gamma_{had})$ | $0.2219 \pm 0.0017$ |
$0.2156^{(1)(1)}_{(1)(1)}$|
$ 0.2159^{(1)(1)}_{(1)(1)}$ | $ 0.2161^{(1)(2)}_{(1)(0)}$ | 
 $0.2167^{(1)(1)}_{(1)(1)}$\|*\\
\|* | | | | | \|*\\
\|*$R(\Gamma_{c\bar{c}}/\Gamma_{had})$ | $0.1543 \pm 0.0074$ | 
$0.1711^{(0)(0)}_{(0)(0)}$|
 $0.1710^{(1)(0)}_{(0)(0)}$|
$ 0.1710^{(0)(1)}_{(1)(0)}$ |
$0.1708^{(1)(0)}_{(0)(1)}$ \|*\\
\|* | | | | | \|*\\
\|*$ A^{0.b}_{FB}$ | $  0.0999\pm 0.0031$|
$0.1003^{(9)( 4)}_{(9)(4)}$ |
$0.1004^{(9)( 3)}_{(8)(4)}$ | $0.1005^{(8)(2)}_{(9)(5)}$ | 
$0.1010^{(10)(4)}_{(8)(3)}$ \|*\\
\|* | | | | | \|*\\
\|*$ A^{0.c}_{FB}$ | $  0.0725\pm 0.0058$|$0.0715^{(7)(3)}_{(7)(3)}$|
$0.0716^{(7)(2)}_{(8)(3)}$| $0.0716^{(7)(2)}_{(8)(3)}$ | 
$0.0720^{(8)(3)}_{(6)(2)}$ \|*\\
\|* | | | | | \|*\\
\|*$ \sin^2\theta^{lepton}_{eff}$ | $0.2325\pm 0.0013 $
| $0.2324^{(1)(1)}_{(2)(1)}$ |
$0.2322^{(2)(0)}_{(1)(2)}$ | 
 $ 0.2322^{(1)(1)}_{(2)(1)}$ | 
 $0.2319^{(1)(0)}_{(2)(2)}$ \|*\\
\|* | | | | | \|*\\
\_
\|*$\chi^2 $ | | 22.6 |20.9| 19.9 | 18.0\|*\\
\_
\|* | | | | | \|*\\
\|*$ \Delta r$ | $  0.0443\pm 0.0102$ 
|$ 0.0390^{(16)(9)}_{(16)(9)}$  |
$0.0396^{(19)(12)}_{(16)(9)}$| $0.0402^{(19)(12)}_{(15)(8 )}$ | 
$0.0423^{(14)(11)}_{(18)(11)}$\|*\\
\_
\EndTable$$
\endinsert

\vglue0.1cm
\leftline{\tenbf 4. Conclusions and Remarks}
 We have examined the results of the minimal $\chi^2$-fits to the precision
measurements of the Z-decay parameters at LEP 
with the aid of a modified ZFITTER program containing the full 
one-loop and dominant two-loop EWRC.
While the result of QBA might appear to be
 in agreement with the 1993 data within $2\sigma$ level of
accuracy~$^{19,20}$, the new world
average value of $M_W$ and updated 1995 LEP data definitely disfavor the QBA,
even though it may seem that the $\chi^2$
value of the QBA is similar to the case of ($m_t$ = 186 GeV, $m_H$ = 1000 GeV)
in Table 1. This is because the origin of the large $\chi^2$ is fundamentally different in two cases: in the latter case, it is mainly due to the large deviation of the predicted
 $R_b$ from the experimental value, while in QBA
it is due to uniform deviation of all $Z$-decay parameters from experiments, i.e., the total $\chi^2$ for the 10 parameters other than  $R_b$ in Table 1 is 10.25 and 17.71 respectively.
Thus the 1995 data support for the non-vanishing electroweak radiative corrections, while the large $\sigma$ of $R_b$
may be the signal for the existence of new physics effect beyond those of 
the SM. In addition, the CDF $m_t$ is a solution of the minimal $\chi^2$-fits 
to the 1995 data with a Higgs scalar mass {\it about} 500 GeV. However this $m_t$ value can be
shifted by as much as 6.4 GeV due to the overall uncertainties in $\alpha_s(M_Z)$ and $\alpha (M_Z)$ for the moment and accordingly $m_H$ 
ranging $320 - 780$ GeV.
Further precision measurement of $M_W$ can provide
a real test of the standard model as it will give a tight constraint for the
needed amount of the EWRC and can provide a profound implication for the mass
of t-quark and Higgs scalar. 
The best-fit solutions within the context of the SM give $M_W$ = 80.29(3) GeV for the CDF range of $m_t$ and $m_H$ = 60 - 1000 GeV. This can be shifted by
another 33 MeV due to the errors of $\alpha_s (M_Z)$ and $\alpha^{-1} (M_Z)$.
 If $M_W$ is determined to within a 30 MeV uncertainty,  $\Delta r$ within the
context of the SM will be tightly constrained to distinguish
 the radiative corrections and the $\chi^2$-fit to the $Z-$decay data with the
 1995 accuracy
 can discriminate the mass range of the t-quark and Higgs scalar within 8 GeV 
 and 200 GeV respectively, providing a crucial test for and even the need of 
 new physics beyond the standard model. If $M_W$ is determined to be larger 
than $80.32$ GeV with
better than a $30$ MeV accuracy by the future precision measurements 
(perhaps reachable at LHC),
 this would be a definite sign for new physics beyond the SM.

The $R_b$ excess at 3.5$\sigma$ and also at a lesser 2.3$\sigma$ $R_c$ deficiency of the $c\bar{c}$ branching ratio  may be the signal of new physics from LEP as we mentioned above. In fact, if we set
$$R_q = R_{q}^{SM} ( 1 + \delta_{q} ), ~~~~~~~ (q = b, c)  \eqno(7)$$
we find $\delta_{b} = 0.0273~~ \pm ~~0.0079$ and $\delta_{c} = -0.0977~~ \pm ~~ 0.0433$,
which are stable with respect to the uncertainties in $\alpha_s$ and $\alpha$ as one can see from Table 1. Other authors~$^{21}$ have noted also that it is impossible to explain $R_b$ and $R_c$ with $\alpha_{s}$ consistent with low energy determinations without introducing new physics corrections to all $Zq\bar{q}$ couplings. There have been many attempts to explain the $R_b$ excess by invoking new physics ingredients beyond the SM. One of the most popular scheme is to use the minimally extended supersymmetric standard model~$^{22}$ which can give additional contributions to $\delta_b$ from the extra supersymmetric particles. In such scheme, one needs either a light higgsino-like chargino and a light supersymmetric partner of the top quark, $\tilde{t}$,
 for small $tan~~ \beta$ or a Higgs pseudoscalar with mass smaller than $M_Z$ when $tan~~ \beta    \gg 1$. Another suggested scenario is to use the extended technicolor model (ETC)~$^{23}$ which has additional technicolor interactions among top quarks and techniq
uarks. Simple ETC models in which the ETC and weak gauge groups commute give a 5 to 20 \% contribution depending on the value of $m_t$ to $\delta_b$ but with an opposite sign. The diagonal techni-neutral ETC bosons can raise $R_b$ but at the expense of introducing large isospin violation, thus causing a unacceptably large oblique parameter $T$. An alternative to the simple ETC model has been proposed by relaxing the commutativity of the ETC and electroweak gauge groups.~$^{23}$ One can achieve the allowed value of $R_b$ by tuning the additional contribution to the $Zb\bar{b}$ vertex from $ZZ^{\prime}$ mixing which are similar in magnitude and opposite in sign to those of the ETC boson exchanges that generate the top quark mass. The phenomenological consequences of this model however are not fully understood.

Finally the indirect bound of the Higgs boson mass has been studied by several authors in MSSM. From a global fit to precision electroweak data, Ellis et al~$^{24}$ estimate $50 < m_H < 124$ GeV for $m_t = 172$ GeV at $36 \%$ probability, while Erler and Langacker~$^{25}$ state $60 < m_H < 150$ GeV leading to  $m_t = 169 \pm 7~~^{+4}_{-3}$ GeV. On the other
hand, we find from our global fit~$^{26}$ that an SM-type $m_H$ is preferred to an MSSM-type Higgs mass within 1$\sigma$ for $m_t = 176 \pm 13$ GeV but if $m_t$ is allowed to vary free, it is difficult to distinguish the two types of the Higgs boson as $60 < m_H < 182$ GeV for $135 < m_t < 162$ GeV within 1$\sigma$.       
\vglue0.4cm
\leftline{\tenbf 5. References}
\itemitem{1.} F. Abe et al., {\it Phys. Rev. Lett.} {\bf 74} (1995) 2626; S. Abachi et al., {\it Phys. Rev. Lett.} {\bf 74} (1995) 2632.

\itemitem{2.} F. Abe {\it et al.}, {\it Phys. Rev. Lett.} {\bf 75} (1995) 11; Fermilab-pub-95/033-E and Fermilab-pub-95/035-1995.

\itemitem{3.} P. Renton, in: Pro. '95LP Symp. (Beijing, August 1995) and OUNP-95-20 (1995).

\itemitem{4.} The LEP Collaborations ALEPH, DELPHI, L3, OPAL and the LEP Electroweak Working Group, CERN Report No. LEPEWWG/95-01; CERN-PPE/95-172. 

\itemitem{5.}  B. A. Kniehl, {\it Int.J.Mod.Phys.} {\bf A10} (1995) 443.

\itemitem{6.}  B. H. Smith and M. B. Voloshin, UMN-TH-1241/94; S. Fanchiotti, B. Kniehl and A. Sirlin, {\it Phys. Rev.} {\bf D 48} (1993) 307; M. Shifman, TPI-MINN-94/42-T.

\itemitem{7.} T. Takeuchi, A. K. Grant and M. P. Worah, FERMILAB-PUB-94/303-T.

\itemitem{8.} T. Hebbeker, Aachen preprint PITHA 91-08 (1991); S. G. Gorishny, A. L. Kataev and S. A. Larin, {\it Phys. Lett.} {\bf B 259} (1991) 144; and L. R. Surguladze and M. A. Samuel, {\it Phys. Rev. Lett.} {\bf 66} (1991) 560.

\itemitem{9.} K. G. Chetyrkin, J. H. Kuhn, and A. Kwiatkowski, TTP94-32 and hep-ph/9503396.

\itemitem{10.} D. Bardin et al., CERN-TH-6443-92 (1992).

\itemitem{11.} See K. Kang and S. K. Kang, {\it Z. Phys.} {\bf C} (in press), SNUTP-94-59 and hep-ph/9509431 for 1994 precision data fits.

\itemitem{12.} Z. Hioki, {\it Mod. Phys. Lett.} {\bf A7} (1992) 1009; K. Kang, in: Proc. 14th Int. Workshop Weak Interactions and Neutrinos (Seoul, 1993), Brown-HET-931 (1993); and K. Kang and S. K. Kang, in: Proc. Workshop on Quantum Infrared Physics (Paris, June 1994), Brown-HET-968 and SNUTP-94-97.

\itemitem{13.} V. A. Novikov et al, {\it Mod. Phys. Lett.} {\bf A8} (1993) 5929.

\itemitem{14.} S. Eidelmann and F. Jegerlehner, {\it Z. Phys.} {\bf C67} (1995) 558.

\itemitem{15.} H. Burkhardt and B. Pietrzyk, {\it Phys. Lett.} {\bf B356} (1995) 398.

\itemitem{16.} W. Hollik, in: {\it Precision Tests of the Standard Model}, ed. P. Langacker (World Scientific Pub., 1993), p. 49; KA-TP-6-1995 and KA-TP-15-1995.

\itemitem{17.} See, for example, W. Hollik, CERN Yellow Book CERN 89-08, Vol.1, p. 45; and K. Kang(Ref. 12).  

\itemitem{18.} M. Consoli and W. Hollik, in {\it Z Physics at LEP 1}, Vol. 1,
eds. G. Altarelli et al., CERN 89-08 (1989).

\itemitem{19.} K. Kang and S. K. Kang, in: Proc. Beyond the Standard Model IV 
(Granlibakken, Lake Tahoe, 1994), Brown HET-979 and SNUTP-94-128 (December, 1994).

\itemitem{20.} K. Hagiwara, S. Matsumoto, D. Haidt and C. S. Kim, {\it Z. Phys.} {\bf C64} (1994) 559; {\bf C68} (1995) 352(E).

\itemitem{21.} See, for example, P. Bamert, McGill/95-64 and NEIP-95-014, and K. Hagiwara, in: Proc. '95LP Symp. (Beijing, August 1995) and KEK-TH-461.

\itemitem{22.} J. D. Wells and G. L. Kane, SLAC-PUB-7038 and hep-ph/9510372. Further References can be found in K. Hagiwara (Ref. 20) and J. Ellis, in this
Proceedings and CERN-TH/95-317.

\itemitem{23.} E. H. Simmons, R. S. Chivukula and J. Terning, in this Proceedings and BU-HEP-95-3x in which earlier References can be found.
\itemitem{24.} J. Ellis (Ref. 21), and  J. Ellis, G. L. Fogli and E. Lisi, CERN-TH/95-202.
\itemitem{25.} J. Erler and P. Langacker, {\it Phys. Rev.} {\bf D52} (1995) 441.
\itemitem{26.} K. Kang and S. K. Kang, Brown-HET-1028 and SNUTP-95-121. 

\medskip
\noindent

\eject

\bye